\documentstyle[aps,multicol]{revtex}

\def\e{\mathop{\rm \mbox{{\Large e}}}\nolimits}

\newcommand{\be}{\begin{equation}}
\newcommand{\ee}{\end{equation}}
\newcommand{\bc}{\begin{center}}
\newcommand{\ec}{\end{center}}
\newcommand{\bea}{\begin{eqnarray}}
\newcommand{\eea}{\end{eqnarray}}
\newcommand{\ba}{\begin{array}}
\newcommand{\ea}{\end{array}}
\newcommand{\nn}{\nonumber}

\def\1{_{1}}
\def\2{_{2}}
\def\y{_{\textstyle{y}}}
\def\z{_{\textstyle{z}}}
\def\dd{^{2}}
\def\bv{\mbox{\boldmath $V$}}
\def\t{\mbox{q}_{\mbox{\small t}}}
\def\ta{\mbox{q}_{\mbox{\small t{\scriptsize 1}}}}
\def\tb{\mbox{q}_{\mbox{\small t{\scriptsize 2}}}}

\begin{document}
\draft

\title{Matching Conditions on Capillary Ripples: Polarization \\
\vspace*{5mm}
Estudio de las condiciones de empalme para las oscilaciones de intercara entre dos fluidos.
Polarizaci\'on.}

\author{Arezky H. Rodr\'{\i}guez \thanks{arezky@ff.oc.uh.cu}, \\
J. Mar\'{\i}n-Antu\~na and H. Rodr\'{\i}guez-Coppola}
\address{Dpto. de F\'{\i}sica Te\'orica, Fac. de F\'{\i}sica, \\
Universidad de la Habana, C. de la Habana 10400, Cuba}

\date{\today}
\maketitle

\begin{abstract}
The matching conditions at the interface between two non-mixed fluids at rest
are obtained directly using the equation of movement of the whole media. This
is a non-usual point of view in hydrodynamics courses and our aim is to fix
ideas about the intrinsic information contained in the matching conditions,
on fluids in this case. Afterward, it is analyzed the polarization of the
normal modes at the interface and it is shown that this information can be
achieved through a physical analysis and reinforced later by the matching
conditions. A detailed analysis of the matching conditions is given to
understand the role that plays the continuity of the stress tensor through
the interface on the physics of the surface particle movement. The main
importance of the viscosity of each medium is deduced.

\vspace*{1cm}

En el presente trabajo se utilizan las ecuaciones de movimiento de todo el
sistema compuesto por dos fluidos no miscibles en reposo relativo para
obtener las condiciones de empalme en la intercara. Este procedimiento es
inusual y el objetivo de hacerlo en esa forma es fijar ideas sobre la
informaci\'on que est\'a contenida en dichas condiciones de empalme, en este
caso en fluidos. Pos\-te\-rior\-men\-te es analizada la polarizaci\'on de los
modos normales en la interface y se demuestra que estas caracter\'{\i}sticas
pueden ser obtenidas directamente de an\'alisis f\'{\i}sico de las
condiciones de empalme.
\end{abstract}


\begin{multicols}{2}

\section{Introduction}

The boundary conditions constitute the key feature in any theory of surface
waves. It is through them that one introduces in the analysis the physical
consequences of specific surface effects, such as local changes of mass or
stresses, thus going beyond the simple case in which one merely matches two
semi-infinite bulk media at an interface (which, in particular, can be a free
surface).

In the last years, increased attention has been paid to the properties of
capillary waves by physicists and chemists \cite{lrl69,l68,tr,hm,y,hb,lt,cd}.
Ripples represent one of the few cases in which the relation between the
dynamical properties of a surface and liquid flows can be predicted
completely. The study of capillary ripples has clarified which properties of
a liquid surface determine the surface's resistance against deformation.

The boundary conditions for the stress at the interface are derived from the
principle that the forces acting upon such an ``interfacial'' element result
not only from viscous stresses in the liquid but also from stresses existing
in the deformed interface. The cause of the difference is that an interface,
unlike a three dimensional liquid, can not enjoy the property of
incompressibility \cite{lrl69}. Work done on an element of liquid is
partially degrade into heat by viscous friction and partially transformed
into kinetic energy which is transmitted to adjoining elements. Work done on
an interfacial element leads, at least partially, to an increment of the
surface potential energy. It is this potential energy of the deformed
interface which enables the whole system, including the interface, to carry
out an oscillatory motion.

In this article, first, it will be shown a non-traditional way to obtain the
matching conditions at the interface between two non-mixed fluid at rest,
considering the equation of motion of the whole media directly. Usually,
university courses do not use this approach and state the boundary conditions
from outside the constitutive equations governing the studied problem. It is
important from a pedagogical point of view to evidence that in fact, the
matching conditions at the interfaces are contained, almost in all cases, in
the equation of movement for the whole system taken as the composition of all
media. The other cases arise when the fluid interfaces have intrinsic
properties not included on the equations of the media in that cases mentioned
before.

This point of view was already used in \cite{gm77,vgm77,vgm79,pvgm81,gmv92}
to elaborate a formalism, based on Surface Green Functions, which  establish
an isomorphism between solids and fluids related to the interface normal
modes. The aim of this paper is to emphasize the fact that this way to
establish the matching conditions allows us to introduce the physical
characteristics related with the boundary conditions in a firmer floating.

\section{Equations for the bulk in a viscous incompressible fluid.}

To achieve a system of equations which describes the oscillatory motion of a
viscous incompressible fluid, the starting point is the linearized
Navier-Stokes equation for a viscous fluid \cite{landau}, which reads:
\be
\label{1.1}
\rho \frac{\partial V_{i}}{\partial t} = \frac{\partial
\tau_{ij}}{\partial x_{j}} + \rho F_{i,\mbox{ext}}
\ee
where $\rho$ is the density of equilibrium, $V_{i}$ the components of the
fluid particle, $F_{i,\mbox{ext}}$ are the external forces and $\tau_{ij}$ is
the stress tensor which, for a viscous incompressible fluid, has the
following form \cite{landau}
\be
\label{1.2}
\tau_{ij} = - p \delta_{ij} + \eta \left( \frac{\partial V_{i}}{\partial
x_{j}} + \frac{\partial V_{j}}{\partial x_{i}} \right)
\ee

Here $p$ are the small variations of pressure, $\eta$ is the viscosity
parameter and $\delta_{ij}$ is the unit matrix elements. The subscripts $i$
and $j$ take the values $y$ and $z$ identically. In (\ref{1.1}) the sum over
repeated subscripts is understood. It had been assumed that the system is
symmetric with respect to the $x$ direction.

Putting (\ref{1.2}) in (\ref{1.1}) it is obtained \bea -\rho \frac{\partial
V\y}{\partial t} + \frac{\partial}{\partial y} \left( -p + 2\eta
\frac{\partial V\y}{\partial y} \right) + \hspace*{15mm} & & \nn \\
\label{1.3} + \; \frac{\partial}{\partial z} \left[ \eta \left(
\frac{\partial V\y}{\partial z} + \frac{\partial V\z}{\partial y} \right)
\right] & = & 0 \\ -\rho \frac{\partial V\z}{\partial t} +
\frac{\partial}{\partial y} \left[ \eta \left( \frac{\partial V\z}{\partial
y} + \frac{\partial V\y}{\partial z} \right) \right] + \hspace*{15mm} & & \nn
\\ \label{1.4} + \; \frac{\partial}{\partial z} \left( -p +
2\eta\frac{\partial V\z}{\partial z} \right) & = & 0 \eea where we have
neglected the external forces.

Also the continuity equation is needed, which expresses that the volume of an
element of the incompressible fluid does not change during the motion. It has
the form \cite{landau}
\be
\label{1.5a}
\frac{\partial V\y}{\partial y} + \frac{\partial V\z}{\partial z} = 0
\ee

The first step is to obtain the equation of motion and its solution of the
each medium taken as infinite. So, the parameters as density and viscosity
are considered constants in the whole medium and it leads to transform
equations (\ref{1.3}) and (\ref{1.4}) to: \bea \label{1.6a} -\rho
\frac{\partial V\y}{\partial t} - \frac{\partial p}{\partial y} + \eta
\nabla\dd V\y & = & 0 \\ \label{1.7a} -\rho \frac{\partial V\z}{\partial t} -
\frac{\partial p}{\partial z} + \eta \nabla\dd V\z & = & 0 \eea where
$\nabla\dd=(\partial\dd/\partial y\dd + \partial\dd/\partial z\dd)$.

The solution of the system (\ref{1.5a}), (\ref{1.6a}) and (\ref{1.7a})
written as a vector field velocity, can be putted as the sum of an
irrotational field (related with the longitudinal mode) and a divergence free
field (related with the transverse modes) \cite{lrl69}, i.e.:
\be
\label{4}
\bv = \bv_{1} + \bv_{2}
\ee
which satisfy:
\bea
\label{1.5}
\nabla \times \bv_{1} & = & 0 \\
\label{1.6}
\nabla\! \cdot\! \bv_{2} & = & 0
\eea

Any irrotational field is characterized by a scalar function, the {\it
``potential'' function} $\varphi(y,z,t)$, such that
\be
\label{12}
\bv_{1} = - \nabla \varphi
\ee
and the divergence free field can be described by a vector function
\cite{landau}, the ``{\it stream}'' or vorticity function $\psi(y,z,t)$, such
that
\be
\label{13}
\bv_{2} = \left(-\frac{\partial \psi}{\partial z},\frac{\partial
\psi}{\partial y} \right)
\ee

The velocity components can thus be written in terms of the potential and
the stream functions, as:
\begin{eqnarray}
\label{14}
V\y & = & - \frac{\partial \varphi}{\partial y} - \frac{\partial
\psi}{\partial z} \\
\label{15}
V\z & = & - \frac{\partial \varphi}{\partial z} + \frac{\partial
\psi}{\partial y}
\end{eqnarray}

Substitution of eqns. (\ref{14}) and (\ref{15}) in the continuity condition
(\ref{1.5a}) gives
\be
\label{16}
\nabla^{2} \varphi = 0
\ee
and substitution into eqns. (\ref{1.6a}) and (\ref{1.7a}) leads to
\bea
\label{17}
\frac{\partial}{\partial y} \left\{ \rho \frac{\partial \varphi}{\partial t}
- p \right\} + \frac{\partial}{\partial z} \left\{ \rho \frac{\partial
\psi}{\partial t} - \eta \nabla\dd\psi \right\} & = & 0 \\
\label{18}
\frac{\partial}{\partial z} \left\{ \rho \frac{\partial \varphi}{\partial t}
- p \right\} - \frac{\partial}{\partial y} \left\{ \rho \frac{\partial
\psi}{\partial t} - \eta \nabla\dd\psi \right\} & = & 0
\eea

Equations (\ref{17}) and (\ref{18}) are simultaneously satisfied if one considers:
\bea
\label{19}
\rho \frac{\partial \varphi}{\partial t} - p & = & C\1 \\
\label{20}
\rho \frac{\partial \psi}{\partial t} - \eta \nabla\dd \psi & = & C\2
\eea

The constants $C\1$ and $C\2$ are obtained from the condition at zero flow,
and it gives raise to $C\1 = p_{o}$ and $C\2 = 0$ where $p_{o}$ is the
reference atmospheric pressure.

The solution of equations (\ref{16}), (\ref{19}) and (\ref{20}) are looking
for as the following \bea \label{23} \varphi(y,z,t) & = & \Phi(z)
\e^{\textstyle{i (\kappa y - \omega t)}} \\ \label{24} \psi(y,z,t) & = &
\Psi(z) \e^{\textstyle{i (\kappa y - \omega t)}} \eea where the parameters
$\kappa$ and $\omega$ are the wavevector and frequency of the wave
respectively.

Substituting eqns. (\ref{23}) and (\ref{24}) in (\ref{16}) and (\ref{20}) gives the
$z$-dependence of the functions $\varphi$ and $\psi$, which satisfy:
\bea
\label{25}
\frac{\mbox{d}^{2} \Phi(z)}{\mbox{d} z^{2}} -
\kappa^{2} \Phi(z) & = & 0 \\ \label{26} \frac{\mbox{d}^{2}
\Psi(z)}{\mbox{d} z^{2}} - \t^{2} \Psi(z) & = & 0
\eea
with $\t^{2} = \kappa^{2} - i \rho \omega/\eta$

Equations (\ref{25}) and (\ref{26}) lead to the solutions:
\bea
\Phi(z) &=& C_{1} \e^{\textstyle{\kappa z}} + C_{2} \e^{\textstyle{-\kappa
z}} \\
\Psi(z) &=& C_{3} \e^{\textstyle{\t z}} + C_{4} \e^{\textstyle{-\t z}}
\eea
and in combination with eqns. (\ref{23}) and (\ref{24}) they give a solution
of the form:
\bea
\label{30}
\varphi(y,z,t) & = & \left( C_{1} \e^{\textstyle{\kappa z}} + C_{2}
\e^{\textstyle{-\kappa z}} \right) \e^{\textstyle{i (\kappa y - \omega
t)}} \\
\label{31}
\psi(y,z,t) & = & \left( C_{3} \e^{\textstyle{\t z}} + C_{4}
\e^{\textstyle{-\t z}} \right) \e^{\textstyle{i (\kappa y - \omega t)}}
\eea
where $C_{1}, C_{2}, C_{3}$ and $C_{4}$ are constants to be determined by
boundary and matching conditions.

Then, the expressions for the varying velocity components and the pressure
are finally obtained by substitution of eqns. (\ref{30}) and (\ref{31}) into
(\ref{14}), (\ref{15}) and (\ref{19}), respectively.

\section{Interface problem: matching conditions.}

Now we will match the two media. We consider a surface which at rest
coincides with the plane $z=0$ and it separates medium $M\1$ at $z<0$ from
medium $M\2$ at $z>0$. Each one is viscous and incompressible.

First of all, any solution has to fulfill the continuity of the velocity
field across the surface according to
\bea
\label{31.5}
V\y^{(1)} & = & V\y^{(2)} \hspace*{1cm} \mbox{at} \; z=0 \\
\label{31.55}
V\z^{(1)} & = & V\z^{(2)} \hspace*{1cm} \mbox{at} \; z=0
\eea

Superscript $1$ denotes medium $M\1$ and superscript $2$ denotes medium
$M\2$. The rest of the matching conditions at the interface are derived from
the system of equations which govern the whole system. This is no usually
done in the normal program courses at the Universities and we consider that
it is important to state that the matching conditions are, almost in all of
the cases, contained in the equation of movement for the whole system taken
as the composition of each one. These are eqns. (\ref{1.3}) and (\ref{1.4})
where the parameters $\eta$ and $\rho$ are constant, but taking different
values on each medium. Integrating these equations through the surface about
$z=0$ from $-\epsilon$ to $+\epsilon$ and later taking $\epsilon \rightarrow
0$, it is obtained from eq. (\ref{1.3}) \bea \left[ \eta_{1} \left(
\frac{\partial V_{\textstyle{y}}^{(1)}}{\partial z} + \frac{\partial
V_{\textstyle{z}}^{(1)}}{\partial y} \right) \right]_{\textstyle{z=0^{-}}} \;
= \hspace*{2cm} & & \nn \\ \label{32}
 = \; \left[ \eta_{2} \left( \frac{\partial V_{\textstyle{y}}^{(2)}}{\partial z} +
 \frac{\partial
V_{\textstyle{z}}^{(2)}}{\partial y} \right) \right]_{\textstyle{z=0^{+}}} & &
\eea
and from eq. (\ref{1.4})
\bea
\left[- p\1 + 2\eta_{1} \frac{\partial V_{\textstyle{z}}^{(1)}}{\partial
z}\right]_{\textstyle{z=0^{-}}} \; = \hspace*{2.5cm} & & \nn \\
\label{33}
= \; \left[ - p\2 + 2\eta_{2} \frac{\partial V_{\textstyle{z}}^{(2)}}{\partial z}
\right]_{\textstyle{z=0^{+}}} - p_{\gamma} & &
\eea
where $p_{\gamma}$ is the jump due to the surface tension according with the
Laplace Law \cite{landau}. The other terms in eqns. (\ref{1.3}) and
(\ref{1.4}) vanish when $\epsilon \rightarrow 0$ because they are continuous
or have a finite jump at $z=0$.

It is easily seen according to eq. (\ref{1.2}) that eqns. (\ref{32}) and
(\ref{33}) are the conditions of the continuity of the stress tensor
components through the interface, as expected.

On the other hand, the boundary conditions of the problem are regularity at
$z \rightarrow \pm \infty$, which leads to eliminate 4 of the 8 constants
appearing in (\ref{30}) and (\ref{31}) for the two media. Using the resulting
functions $\varphi$ and $\psi$ for each medium in conditions
(\ref{31.5})-(\ref{33}) the following system is met:
\be
\label{n5}
\left\Vert \begin{array}{cccc}
 -i \kappa & i \kappa & - \ta & - \tb \\
 -1 & -1 & i & -i \\
-2 i \kappa^{2} \eta_{1} & -2 i \kappa^{2} \eta_{2} & -\eta_{1} Q_{t1} &
\eta_{2} Q_{t2} \\ d_{41} & \eta\2 Q_{t2} & d_{43} & 2 i \kappa \eta_{2} \tb
\\
\end{array} \right\Vert
\left\Vert \begin{array}{c}
 C_{1} \\
 C_{2} \\
 C_{3} \\
 C_{4} \\
\end{array} \right\Vert
= 0
\ee
to determine the remaining constants, with the definitions $d_{41}= -
\eta\1 Q_{t1} - i \Upsilon$, $d_{43} = 2 i \kappa
\eta_{1} \ta - \Upsilon$,  $\Upsilon =
\gamma \kappa^{3}/\omega$, $Q_{t1} = 2\kappa\dd - i \rho\1 \omega/\eta\1$ and
$Q_{t2} = 2\kappa\dd - i\rho\2 \omega/\eta\2$.

In deducing the expression in system (\ref{n5}) which comes from (\ref{33})
it was considered that
\be
\label{45} p_{\gamma} = - \gamma \frac{\partial^{2} z}{\partial y^{2}} \ee on
the surface, but as we are dealing with the velocity field, then it is
obtained that:
\be
\label{46}
\frac{\partial p_{\gamma}}{\partial t} = - \gamma \frac{\partial^{2}
V\z}{\partial y^{2}}
\ee
which finally leads to:
\be
\label{47}
p_{\gamma} = - \frac{\gamma \kappa^{2}}{i \omega} \left( -\kappa
C_{1} + i \kappa C_{3} \right) \e^{\textstyle{i(\kappa y - \omega
t)}}
\ee
according to the substitution $\partial/\partial t \rightarrow - i \omega$ and
$\partial/\partial y
\rightarrow - \kappa^{2} $

The vanishing of the determinant of (\ref{n5}) leads to the equation for the
dispersion relation of the existing surface modes. It can be written as:
\begin{eqnarray}
\omega^{2} \left[(\rho_{1} + \rho_{2})(\rho_{1} \tb + \rho_{2} \ta)
- \kappa (\rho_{1} - \rho_{2})^{2} \right] \; & + &  \nn
\\
+ \; \gamma \kappa^{3} \left[ \rho_{1} (\kappa - \tb) + \rho_{2} (\kappa - \ta)
\right] \; & + &  \nn
\\
+ \; 4 \kappa^{3} (\eta_{2} - \eta_{1})^{2} (\kappa - \ta)(\kappa - \tb) & + & \nn \\
\label{50}
+ \; 4 i \kappa^{2} \omega (\eta_{2} - \eta_{1}) (\rho_{1} \kappa -
\rho_{2} \kappa - \rho_{1} \tb + \rho_{2} \ta) & = & 0
\end{eqnarray}

This is the dispersion relation for capillary waves for the interface of two
viscous incompressible fluids.

In order to obtain the constants $C_{1}, C_{2}, C_{3}$ and $C_{4}$, an
initial stimulus is needed according to an initial value problem \cite{dr}
but this method, simple at the beginning, becomes rather complicated later
and it is not good for a quite general study.

As was said on the introduction, the aim of this paper is not to get inside
the dispersion relation of the normal modes at the interface of two viscous
fluids at rest. For a better study of this subject we recommend paper
\cite{nosotros}. We get here to show a way of solution also different from
the usually taken as only the velocity vector as a function of an ''{\it
stream  function}''. From now on, we will put our attention on the matching
conditions and we will show that more that a mathematical information of the
matching can be found on it, but also the physics of the polarization can be
deduced and how the interface moves in its oscillation.

From eqns. (\ref{32}) and (\ref{33}) it can be obtained more information about the
polarization of the modes on the interface. This will be done in the next section.

\section{Matching conditions and polarization.}

There are two possible modes on the fluid: one in which the fluid particle
moves in the direction of the wave propagation called {\it longitudinal} with
notation L($V\y$,$0$) and another {\it transverse} to the direction of the
wave propagation and {\it normal} to the interface with notation
TN($0$,$V\z$).

The longitudinal mode is related to the fluid compressibility because this
motion of the fluid particles is only possible when its volume changes
\cite{landau}. This analysis also holds when the mode is on the interface,
but this does not mean that there are no longitudinal modes of oscillation on
the interface when the fluids involved are incompressible. It had been shown
in \cite{lrl69} that the interface, when oscillating, must be considered as a
compressible one, because precisely its change in area is responsable for the
increasing of its potential energy and therefore, for its oscillation. It is
important to state that this is a fundamental argument to understand the
movement of any interface in hydrodynamics.

Nevertheless, now it can be shown that in spite of the compressibility of the
interface, there does not exist pure longitudinal mode if we are dealing with
incompressible media. Let us demonstrate this.

If a point $y_{o}$ on the interface is considered moving with velocity, say
$V_{\textstyle{yo}}^{S}$ in the $y$-axis direction, then, according to the
continuity of velocity, the point ($y_{o}$,$-\epsilon$) in $M\1$ and the
point ($y_{o}$,$+\epsilon$) in $M\2$ must have the same velocity
$V_{\textstyle{yo}}^{S}$ if $\epsilon$ is small enough. As the interface is
compressible, at the point $y_{1}$ near enough $y_{o}$ the velocity can be,
for instance, $V_{\textstyle{y1}}^{S}$ different in general from
$V_{\textstyle{yo}}^{S}$ but as the media are incompressible, at the point
($y_{1}$,$-\epsilon$) and ($y_{1}$,$+\epsilon$) the velocity must be
$V_{\textstyle{yo}}^{S}$. See Fig. 1. This is not in conformity with the
continuity of the velocity through the surface and hence the pure
longitudinal mode is not possible and only the TN mode seems to be valid when
the media are incompressible.

After these considerations during the above demonstration, the student can
keep the idea that the interface oscillations can only occur in the $z$-axis.
This is the accurate moment to show to the student that things not always are
as they apparently seem to be, because that assumption does not take into
account the different properties of each medium, whose response depends on
its fundamental parameters, as density and viscosity, which are different for
each medium. Hence, it is evident that it must be analyzed, precisely, the
interface matching conditions.

It is useful to compare and to support the previous qualitative analysis with
a quantitative and more profound one regarding the interface matching
conditions.

Recalling carefully eqns. (\ref{32}) and (\ref{33}) and supposing that such a
wave propagates in $y$ direction with movement only in $z$ direction (TN
mode), then $V\y^{(1)}=V\y^{(2)}=0$ and eq. (\ref{32}) becomes
\be
\label{51}
\left. \eta\1 \frac{\partial V\z^{(1)}}{\partial y}
\right\vert_{\textstyle{z=0^{-}}} = \left. \eta\2 \frac{\partial
V\z^{(2)}}{\partial y} \right\vert_{\textstyle{z=0^{+}}}
\ee

It is known that $V\z$ is continuous along the interface for all points.
Then, the derivative with respect to $y$ is also the same in both hands of
(\ref{51}) and this expression only holds if $\eta\1=\eta\2$, i.e., if the
media have the same viscosity. It does not mean for the interface to
disappear because the density of each medium can be different. Then, if the
viscosities of the media have not the same value, the velocity component
$V\y$ along the interface must be different from zero to compensate the
inequality (\ref{51}) and to fulfill the continuity of the stress tensor in
the $y$ direction given by eq. (\ref{32}) yielding to a component of movement
along $y$ direction. This mode will be called {\it Sagittal} mode or
S$(V\y,V\z)$.

The above analysis was done for the general case. Now we are able to take the
particular case in which one of the media is vacuum, for instance, $M\2$,
with $\eta\2=0$. Then, condition (\ref{32}) becomes
\be
\label{52} \left[ \eta\1 \left( \frac{\partial V\y^{(1)}}{\partial z} +
\frac{\partial V\z^{(1)}}{\partial y} \right) \right]_{\textstyle{z=0^{-}}} =
0 \ee and it can be seen that also $V\y$ must be non zero on the surface to
hold eq. (\ref{52}) with the corresponding {\it Sagittal} polarization
movement.

With this analysis on the conditions of stress component continuity in $y$
direction along the interface, it can be seen that if the two media are
viscous (at least one of them), the fluid particle of the interface moves in
a {\it Sagittal} mode which combines movement in both directions: along the
wave propagation in $y$ direction, and normal to the interface in $z$
direction.

Then, it is qualitatively clear that the viscosity of each media plays a
fundamental role in the coupling of modes even for incompressible fluids. In
spite of that, it could be a mistake to say that the modes decouple if the
viscosities are equal. It should not be forgotten that eq. (\ref{33}) is also
important in the characterization of the interface particle behaviour and it
includes the pressure on each side of the surface. According to eq.
(\ref{19}), the pressure is associated with the longitudinal mode and the
inertial effect of the fluid particle according with the density of the
media. Then, it contains the information of each components of the velocity
and also of the density and according to eq. (\ref{30}) the pressure has a
jump through the interface. This result, in combination with the analysis of
eq. (\ref{32}) make difficult to understand the role played by the densities
of each media on the surface polarization movement, and it can not be reached
from this only analysis. This point is still a matter of investigation.

\section{Conclusions}

The present work is an attempt to give an example, using the hydrodynamics,
of how the study of the interface matching conditions allows us to make a
plentiful and rich in details discussion. Moreover, of how the interface
matching conditions content a sufficient information to conclude that the
interface oscillation must be with a {\it Sagittal} mode and not with neither
a pure longitudinal, nor a pure transversal one. This movement has been shown
to be close related to the physical properties of the media such as
viscosities and that fact allows us to establish rigorously that those are
the parameter which characterize the interface movement and the response of
each medium to an stimulus coming from the other one.

It was seen that viscosity is the main parameter in the coupling of the two
modes to achieve a Sagittal one, nevertheless within the framework of this
formalism it is difficult to determine the role of viscosity and of the
density ratios in the coupling of modes. This aspect needs further
investigation.

\begin{figure}
\caption{Relation of velocities on the interface between two viscous fluids.}
\end{figure}

\end{multicols}

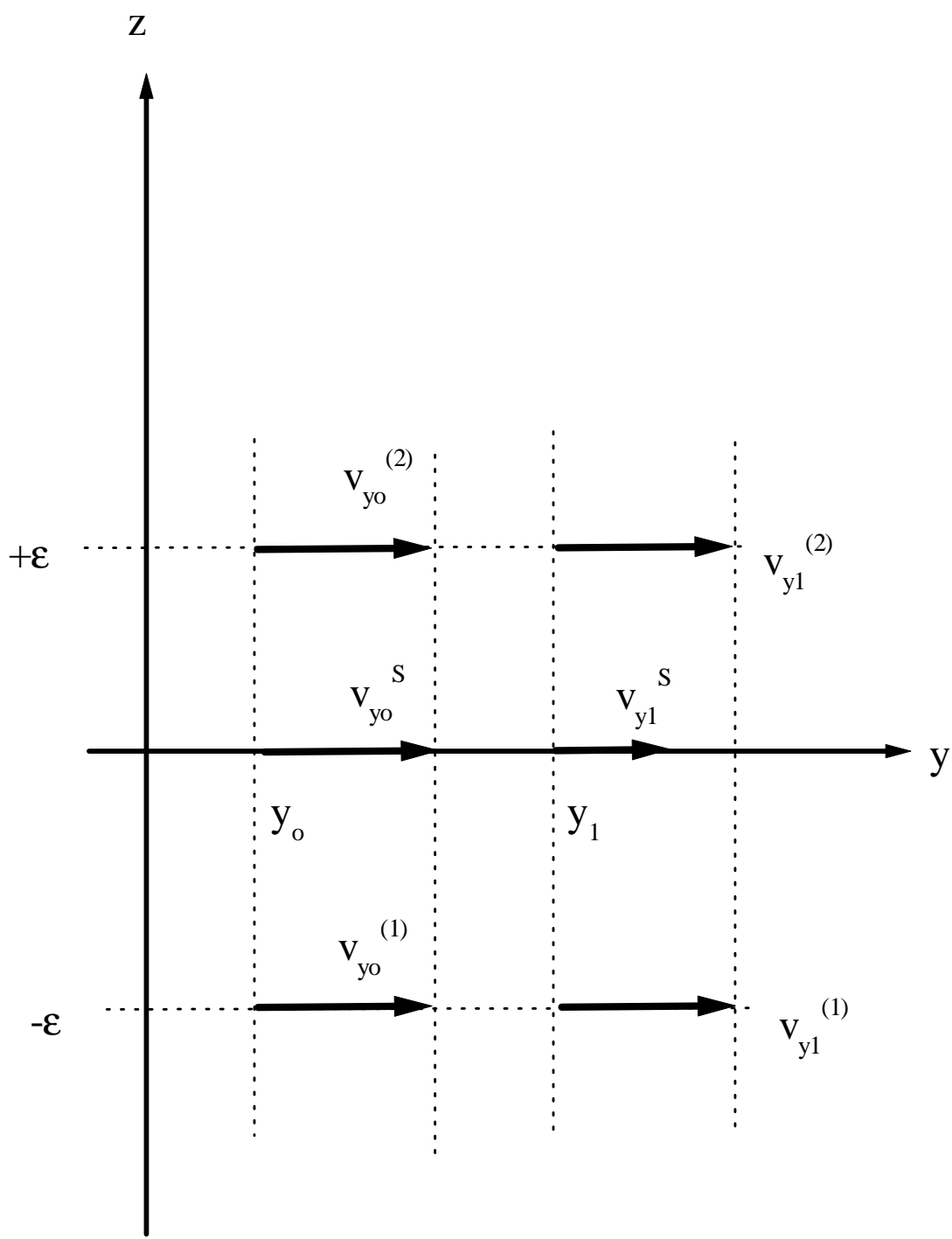
\end{document}